\documentclass{aa}  
\usepackage{graphicx}
\usepackage{txfonts}
\usepackage[varvw]{newtxmath}	
\usepackage{geometry}
\usepackage{xcolor}

\usepackage{hyperref}
\hypersetup{
    colorlinks=true,
    linkcolor=blue,
    filecolor=magenta,      
    urlcolor=cyan,
    citecolor=blue,
    pdftitle={Overleaf Example},
    pdfpagemode=FullScreen,
    }

\makeatletter
\renewcommand*\aa@pageof{, page \thepage{} of \pageref*{LastPage}}
\makeatother
\begin{document}

    \title{Multi-thermal dynamics and transverse oscillations of solar spicules revealed by coordinated SST, IRIS, and SDO observations}

       \author{Ravi Chaurasiya\inst{1,2}\fnmsep\thanks{Corresponding author}
    	\and Tiago M. D. Pereira\inst{3,4}\fnmsep
    	\and A. Raja Bayanna\inst{1}}
    \institute{Udaipur Solar Observatory, Physical Research Laboratory, Udaipur-313001, India
    	\and Indian Institute of Technology, Gandhinagar, Gujarat-382355, India
   	\and Rosseland Centre for Solar Physics, University of Oslo, PO Box 1029 Blindern, 0315 Oslo, Norway
   	\and Institute of Theoretical Astrophysics, University of Oslo, PO Box 1029, Blindern, 0315 Oslo, Norway }

  \date{Received: XXXX / Accepted: XXXX}

 
\abstract
{Solar spicules are highly dynamic chromospheric jets that play an important role in the mass and energy balance of the solar atmosphere, though their connection to the transition region and corona remains unclear.}
{We investigate the dynamical and multi-thermal properties of spicules, their connection to higher atmospheric layers, and their transverse oscillations and associated energy flux.}
{We analyse coordinated high-resolution observations from the Swedish 1-m Solar Telescope (SST) in H$\alpha$, the Interface Region Imaging Spectrograph (IRIS) in \ion{Si}{iv} 1400~\AA, and the Solar Dynamics Observatory (SDO)/Atmospheric Imaging Assembly (AIA) coronal channels. Space–time diagrams, spectral analysis, and wavelet techniques are used to study temporal evolution, Doppler velocities, and oscillatory properties. Transverse displacements of spicules are tracked to estimate wave properties and energy flux.}
{Space–time analysis reveals a clear correspondence between chromospheric spicules and coronal emission in AIA 171~\AA, evolving coherently with spicule extension. Doppler velocities from H$\alpha$ and \ion{Si}{iv} show opposite signs, indicating multi-thermal plasma flows. Wavelet analysis reveals frequently dominant $\sim$3-minute oscillations, along with high-frequency transverse oscillations (65--270 s) with velocity amplitudes of 3.3--9.9 km s$^{-1}$ and a mean energy flux of $(2.14 \pm 0.78)\times10^{3}$ W m$^{-2}$.}
{These results demonstrate that spicules are multithermal, dynamic structures connected to the transition region and corona, and that transverse waves carry substantial energy, highlighting their role in coronal heating.}


    \titlerunning{Spicules in the Solar Atmosphere }
    \authorrunning{Chaurasiya, Pereira, \& Bayanna}

   \maketitle
%

\section{Introduction}

The solar chromosphere is a highly dynamic and structured layer that plays a crucial role in mediating the transfer of mass and energy between the photosphere and the corona \citep{Pereira.19,Carlsson.et.al.19}. This complex and highly structured region is shaped by the intricate interplay between magnetic fields, plasma dynamics, and radiative processes. As a result, it hosts a wide variety of dynamic phenomena, including numerous jet-like structures such as spicules \citep{Beckers.68,DePontieu.et.al.04,DePontieu.et.al.07,Pereira.et.al.12,Pereira.et.al.14,Pereira.et.al.16,Chaurasiya.et.al.24}, as well as pervasive wave activity that propagates through different atmospheric layers \citep{Zaqarashvili&Erdelyi.09,Jess.et.al.23,Chaurasiya&Bayanna.25,Chaurasiya.et.al.25}. 

Among the most prominent chromospheric features are spicules, which are ubiquitously observed at the solar limb in chromospheric spectral lines such as H$\alpha$ and \ion{Ca}{ii} H. Spicules exhibit rapid evolution, with typical lifetimes of a few minutes, lengths of several megametres, and upward velocities ranging from a few to tens of km s$^{-1}$ \citep{Beckers.68,DePontieu.et.al.07,Pereira.et.al.12}. Due to their ubiquity and dynamic nature, spicules have long been considered potential contributors to coronal mass supply and heating.

Recent high-resolution observations have revealed that spicules exhibit complex multi-thermal behaviour, with signatures extending from chromospheric to transition region temperatures. Observations from the Interface Region Imaging Spectrograph (IRIS) have shown that spicular plasma can reach transition region temperatures, suggesting that spicules may play an important role in supplying heated plasma to higher atmospheric layers \citep{Pereira.et.al.14,Skogsrud.et.al.15,Chaurasiya.et.al.24}. Furthermore, coordinated observations with the Solar Dynamics Observatory (SDO) have revealed coronal counterparts associated with spicular activity, supporting the idea that spicules may contribute to coronal dynamics and energy balance \citep{DePontieu.et.al.11,Henriques.et.al.16,Samanta.et.al.19,Chaurasiya.et.al.24,Chaurasiya.et.al.26}. However, the exact physical relation between chromospheric spicules and coronal emission remains a subject of ongoing debate.

In addition to their thermal evolution, spicules exhibit a wide range of wave motions, including Alfven waves, torsional motions, and transverse oscillations \citep{DePontieu.et.al.07b,DePontieu.et.al.12,Okamoto&DePontieu.11,Shetye.et.al.21,Bate.et.al.22}. These waves are of particular interest because they can transport significant energy into the upper solar atmosphere and may contribute to atmospheric heating. Previous studies have reported oscillations with periods ranging from tens of seconds to several minutes, including the ubiquitous three-minute chromospheric oscillations, which are thought to be associated with upward-propagating magneto-acoustic waves originating in the lower atmosphere \citep{Jess.et.al.15,Jess.et.al.23}. However, most studies have focused on vertical velocity oscillations rather than the horizontal oscillations.

Despite significant progress, several key questions remain unresolved. In particular, the temporal and spatial relation between chromospheric spicules and coronal emission, the multi-thermal velocity structure of spicules, and the role of oscillations in driving spicule dynamics are not yet fully understood. Coordinated multi-wavelength observations combining chromospheric, transition region, and coronal diagnostics provide a powerful means to address these questions.

In this study, we analyse coordinated high-resolution observations of spicules obtained with the Swedish 1-m Solar Telescope (SST), the Interface Region Imaging Spectrograph (IRIS), and the Solar Dynamics Observatory (SDO). These observations allow us to simultaneously investigate the chromospheric, transition region, and coronal response associated with spicule evolution. Using space–time analysis, spectroscopic diagnostics, and wavelet techniques, we examine the multi-thermal dynamics, velocity structure, and transverse oscillatory behaviour of spicules, and their connection to coronal emission.

The rest of the paper is structured as follows. Section~\ref{Sec2} describes the observations used in this study. Section~\ref{Sec3} presents the results, including the multiwavelength evolution of the solar atmosphere and the wave analysis. The results are discussed in Section~\ref{Sec4} followed by the conclusion in Section \ref{Sec5}.

\section{Observations} \label{Sec2}

We analysed coordinated datasets of a quiet-Sun (QS) region near the north pole of the Sun, observed on 17 June 2014 between 10:20 UT and 11:15 UT. The heliocentric coordinates of the target region were ($24\arcsec$, $939\arcsec$) giving rise to $\mu = \cos\theta = 0.207$, where $\theta$ is the heliocentric angle between the observer's line-of-sight and the local solar vertical. The observed region was predominantly quiet, without any major active region or plage. Further information regarding the data from various instrumentation facilities is provided below.

\begin{figure}
	\includegraphics[width=85mm]{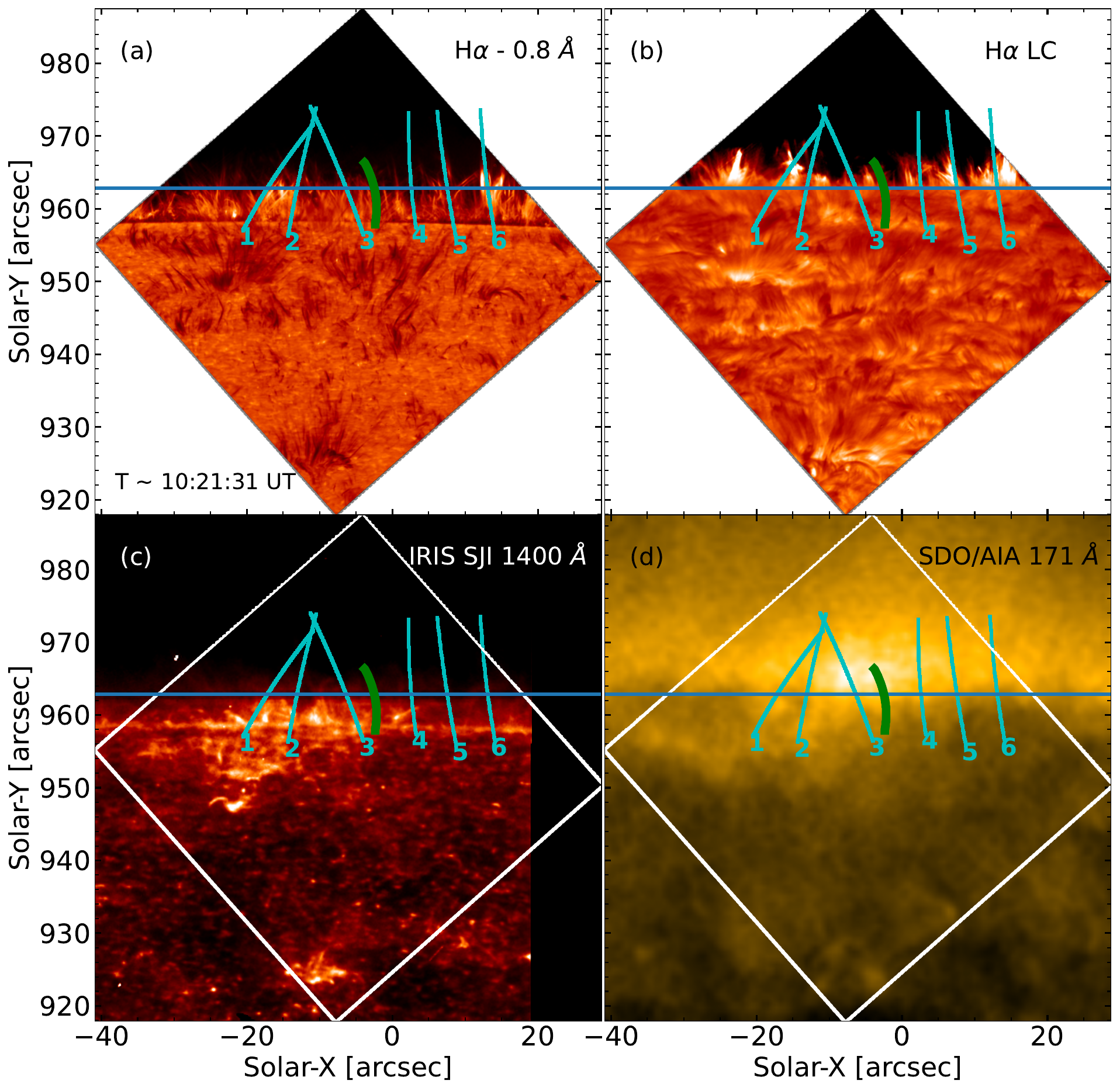}
	\caption{Overview of coordinated multiwavelength observations sampling the chromosphere, transition region, and corona. Panels (a) and (b) show the SST H$\alpha$ $-0.8$~\AA\ wing and H$\alpha$ line core images, respectively, representing the lower and middle chromosphere. Panel (c) displays the IRIS SJI 1400~\AA\ image, which samples the transition region, while panel (d) shows the SDO/AIA 171~\AA\ image, representing the million-degree corona. The blue horizontal line indicates the location of the IRIS spectrograph slit. The green line marks the positions of the virtual slit used to construct the space–time diagrams as shown in Figure \ref{fig:parabolic_path}.The six cyan lines mark the positions of the selected virtual slits, which are used to construct multi-thermal space–time diagrams for investigating the temporal and spatial evolution of structures across different atmospheric layers. The white diamond outlines the common field of view of the SST observations, overplotted on the IRIS and AIA panels for reference.}
	\label{overview}
\end{figure}

\begin{figure}
	\includegraphics[width=90mm]{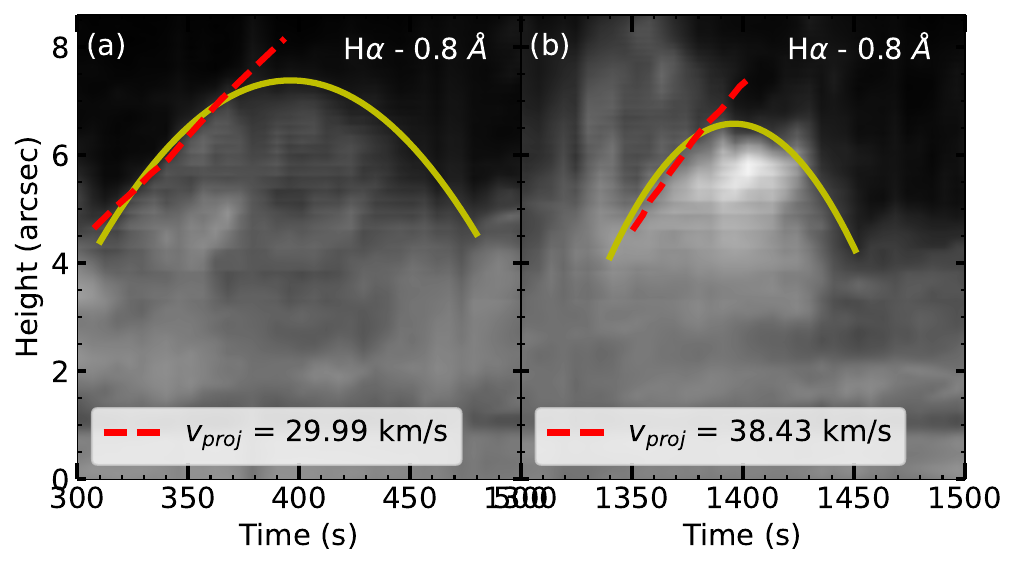}
	\caption{Representative examples of limb spicules exhibiting parabolic trajectories in the H$\alpha$ - 0.8 \AA~ space--time diagrams for the green slit shown in Figure \ref{overview}. The yellow curves indicate parabolic fits tracing the temporal evolution of the spicule apex, while the red dashed lines represent linear fits to the initial upward phase, used to estimate the projected upward velocity (v$_{proj}$).}
	\label{fig:parabolic_path}
\end{figure}

\subsection{Swedish 1-m Solar Telescope}

The SST dataset consists of imaging spectroscopic observations of the H$\alpha$ spectral line acquired with the CRisp Imaging SpectroPolarimeter (CRISP; \citealt{Scharmer.et.al.08}) mounted on the SST. The observations cover a duration of approximately 55 minutes, with a temporal cadence of 5.5~s. The seeing conditions were excellent during this period. The field of view is about $61\arcsec \times 61\arcsec$, and the spatial sampling is $0\farcs058$ per pixel.

The H$\alpha$ line was sampled at 25 wavelength positions spanning a spectral range of $\pm1.2$~\AA\ around the line centre. The data were processed using the CRISPRED reduction pipeline \citep{delaCruzRodriguez.15}, which includes standard calibration procedures such as dark subtraction, flat-field correction, and image restoration. Image reconstruction was performed using the Multi-Object Multi-Frame Blind Deconvolution (MOMFBD) technique \citep{VanNoort.et.al.05} to improve the spatial resolution. Additional details about this SST observation and reduction procedures can be found in \citet{Pereira.et.al.16}.

\subsection{Interface Region Imaging Spectrograph (IRIS)}

We used coordinated level~2 data from the Interface Region Imaging Spectrograph (IRIS; \citealt{DePontieu.et.al.14}) obtained under OBSID 3820259453 as part of the SST--IRIS joint observing campaign which consisted of both slit-jaw images(SJIs) and spectroscopic observations. The IRIS observations cover the interval from 07:30 UT to 11:00 UT on 17 June 2014, overlapping with most of the SST observing period but not covering the final 15 minutes. The level~2 data were processed using the standard IRIS reduction pipeline (\texttt{iris\_prep}, version~2.65), which includes dark current subtraction, flat-field correction, geometrical correction, and wavelength calibration. 

SJIs were acquired in the \ion{Si} {iv} 1400~\AA\ and \ion{Mg} {ii} k 2796~\AA\ channels, both with a spatial sampling of $0\farcs166$ per pixel. The average temporal cadence of the SJIs was 18.82~s. These channels sample plasma formed in the transition region and upper chromosphere, respectively. In addition to the imaging observations, IRIS performed spectroscopic measurements in a Large sit-and-stare mode with an effective cadence of 9.2~s. The spectral windows include chromospheric and transition region lines such as \ion{C} {ii} 1336~\AA, \ion{Si} {iv} 1394~\AA\ and 1403~\AA, and \ion{Mg} {ii} k 2796~\AA. In the present study, we focus on the \ion{Si} {iv} 1394~\AA\ line, which has a spectral sampling of approximately 25.44~m\AA. The absolute wavelength calibration provided by the level~2 pipeline is based on neutral reference lines formed in the lower solar atmosphere, where intrinsic velocities are typically small \citep{DePontieu.et.al.14}.

For co-alignment purposes, the IRIS SJI images were resampled to match the spatial sampling of the SST observations using bilinear interpolation. The alignment was refined by cross-correlating nearly simultaneous IRIS SJI 2796~\AA\ images with the corresponding SST wideband images.

\begin{figure*}
	\centering
	\begin{tabular}{cc}
		\includegraphics[width=85mm]{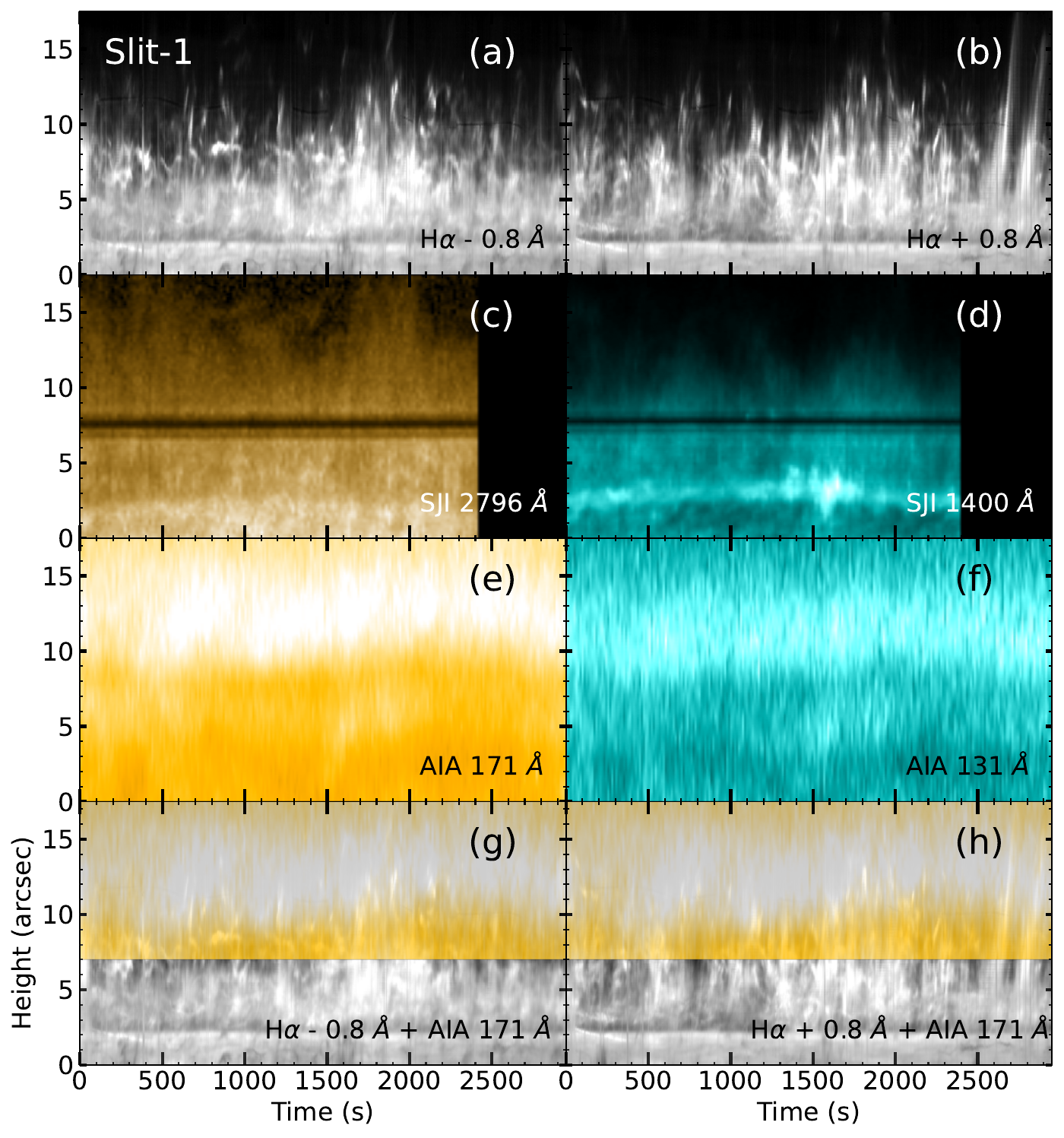} &
		\includegraphics[width=85mm]{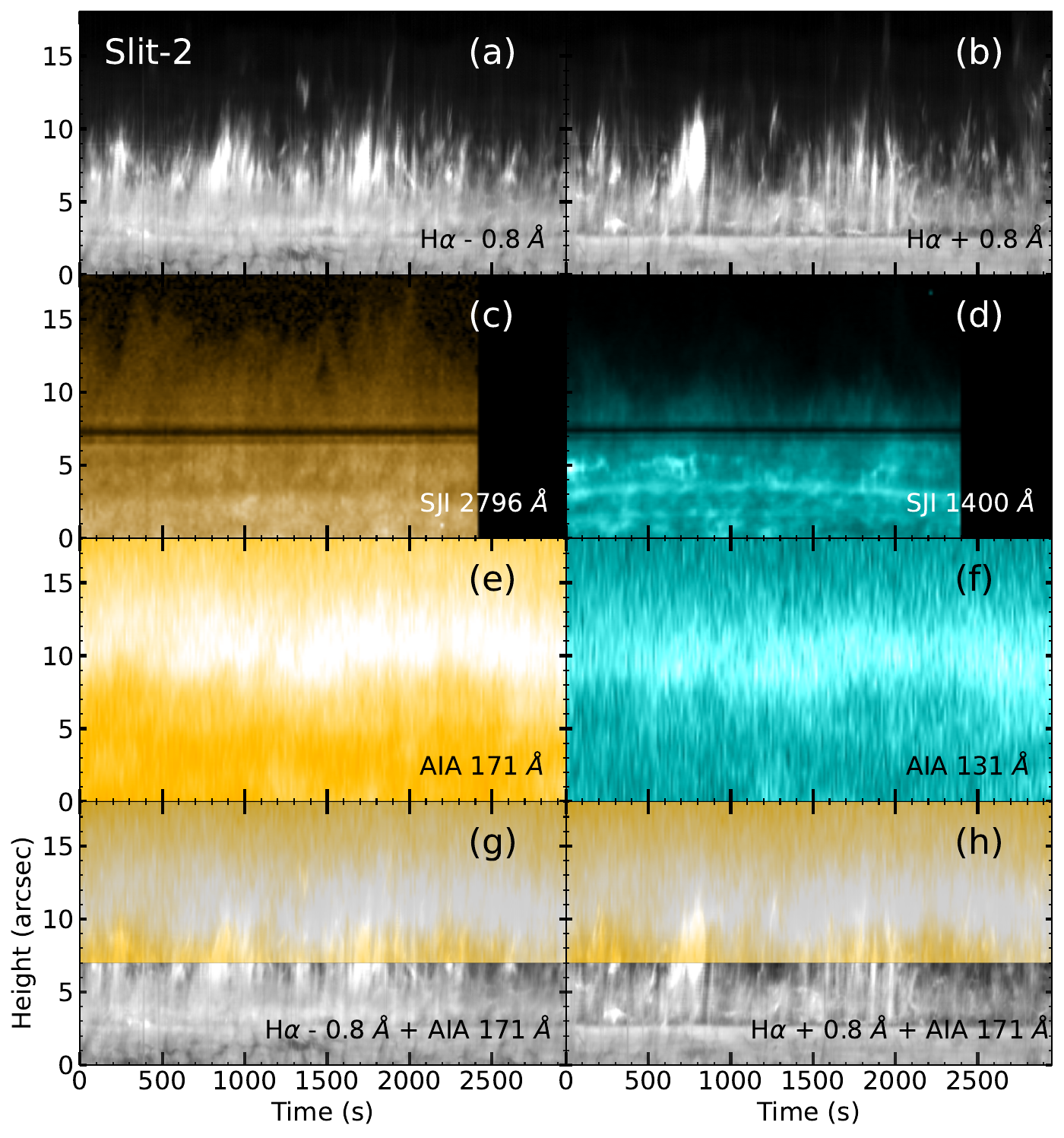} \\
		\includegraphics[width=85mm]{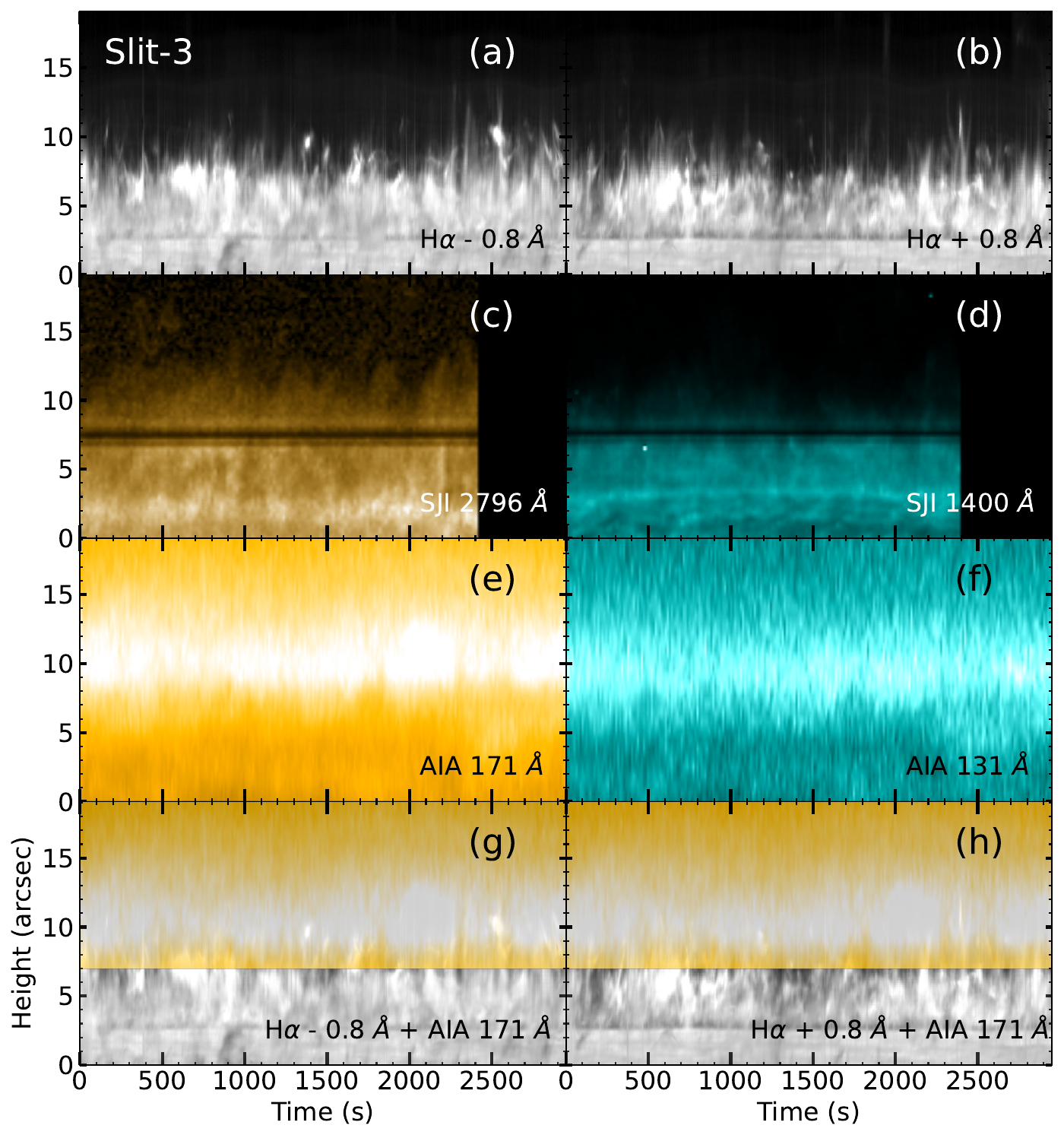} &
		\includegraphics[width=85mm]{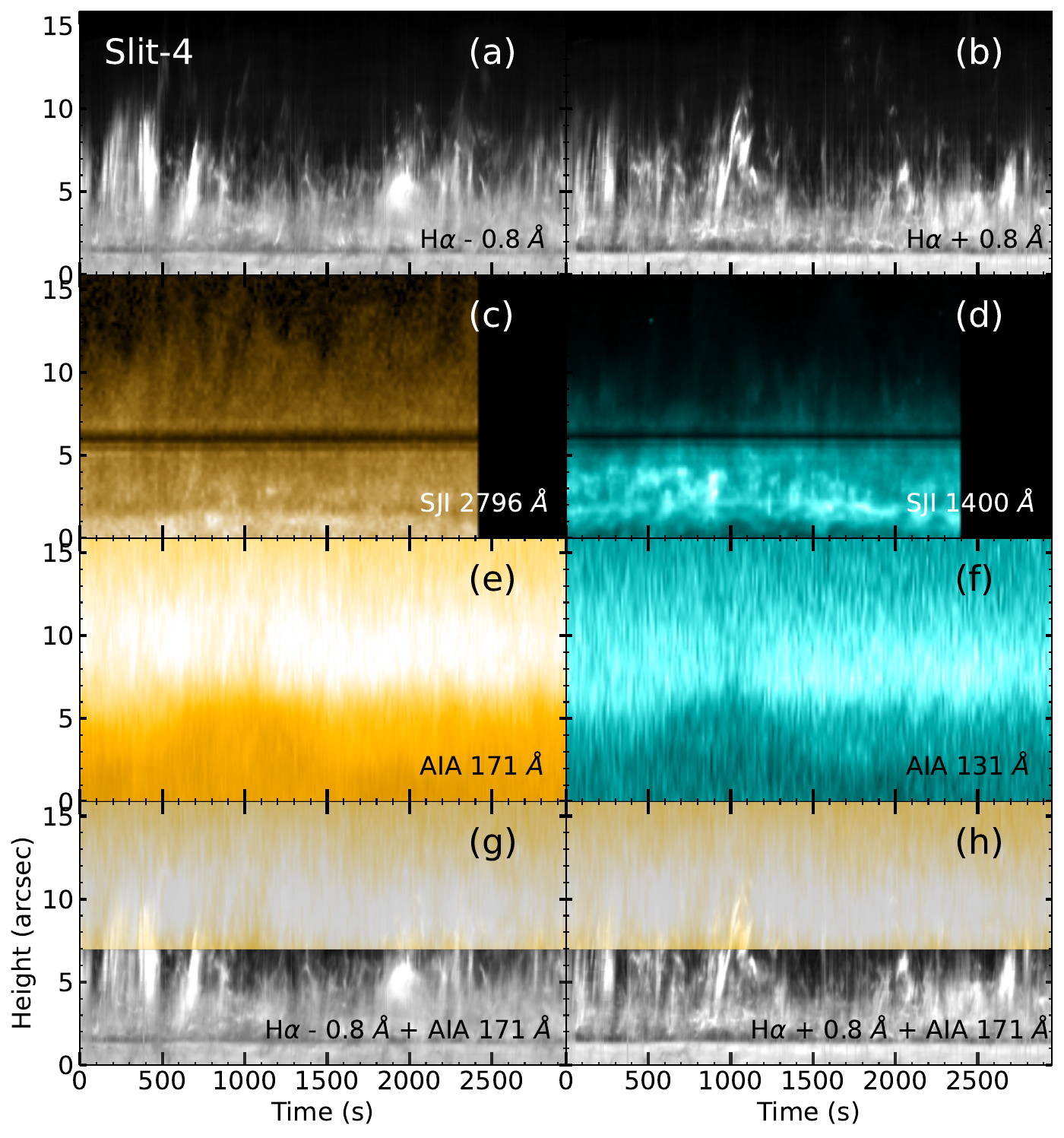} \\
	\end{tabular}
	
	\caption{
		Multi-thermal space--time maps constructed along four different slits (see Figure~\ref{overview}) using coordinated SST, IRIS, and SDO observations, illustrating the temporal and spatial evolution of plasma across different layers of the solar atmosphere. In each example, the top row shows the SST H$\alpha$ $-0.8$~\AA\ (left) and H$\alpha$ $+0.8$~\AA\ (right) wing maps, sampling the lower chromosphere and revealing numerous upward-propagating jet-like structures. The second row presents the IRIS SJI 2796~\AA\ (left) and SJI 1400~\AA\ (right) maps, representing the upper chromosphere and transition region, respectively. The third row shows the corresponding coronal emission observed in the SDO/AIA 171~\AA\ (left) and 131~\AA\ (right) channels. The bottom row displays composite maps formed by overlaying H$\alpha$ $-0.8$~\AA\ with AIA 171~\AA\ (left) and H$\alpha$ $+0.8$~\AA\ with AIA 171~\AA\ (right), highlighting the temporal and spatial correspondence between chromospheric jets and their coronal counterparts. The horizontal axis represents time, while the vertical axis shows the projected height along the slit.}
	
	\label{Space_time_map_slt_1_2_3_4}
\end{figure*}

\begin{figure*}
\includegraphics[width=90mm]{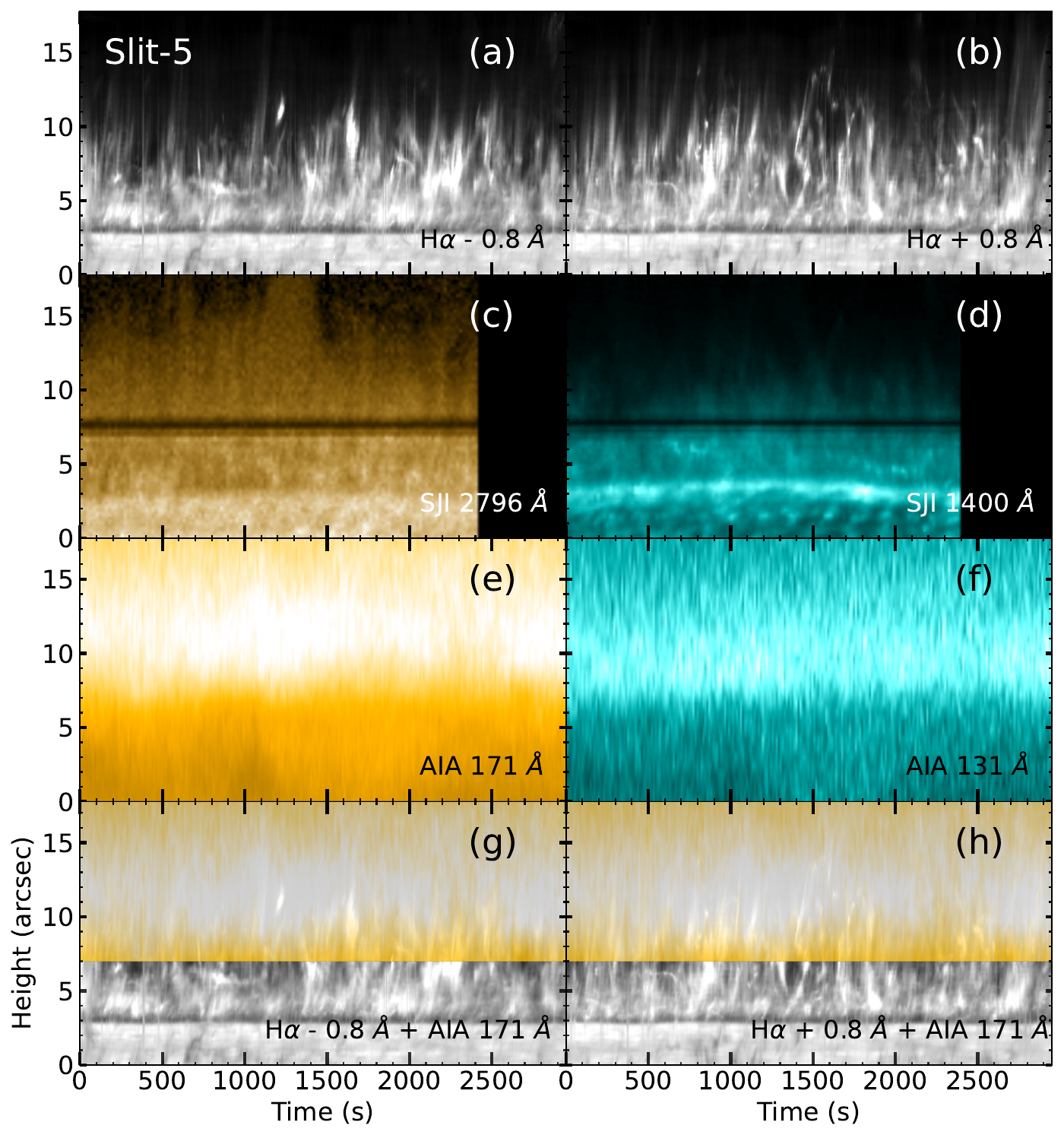}
\includegraphics[width=90mm]{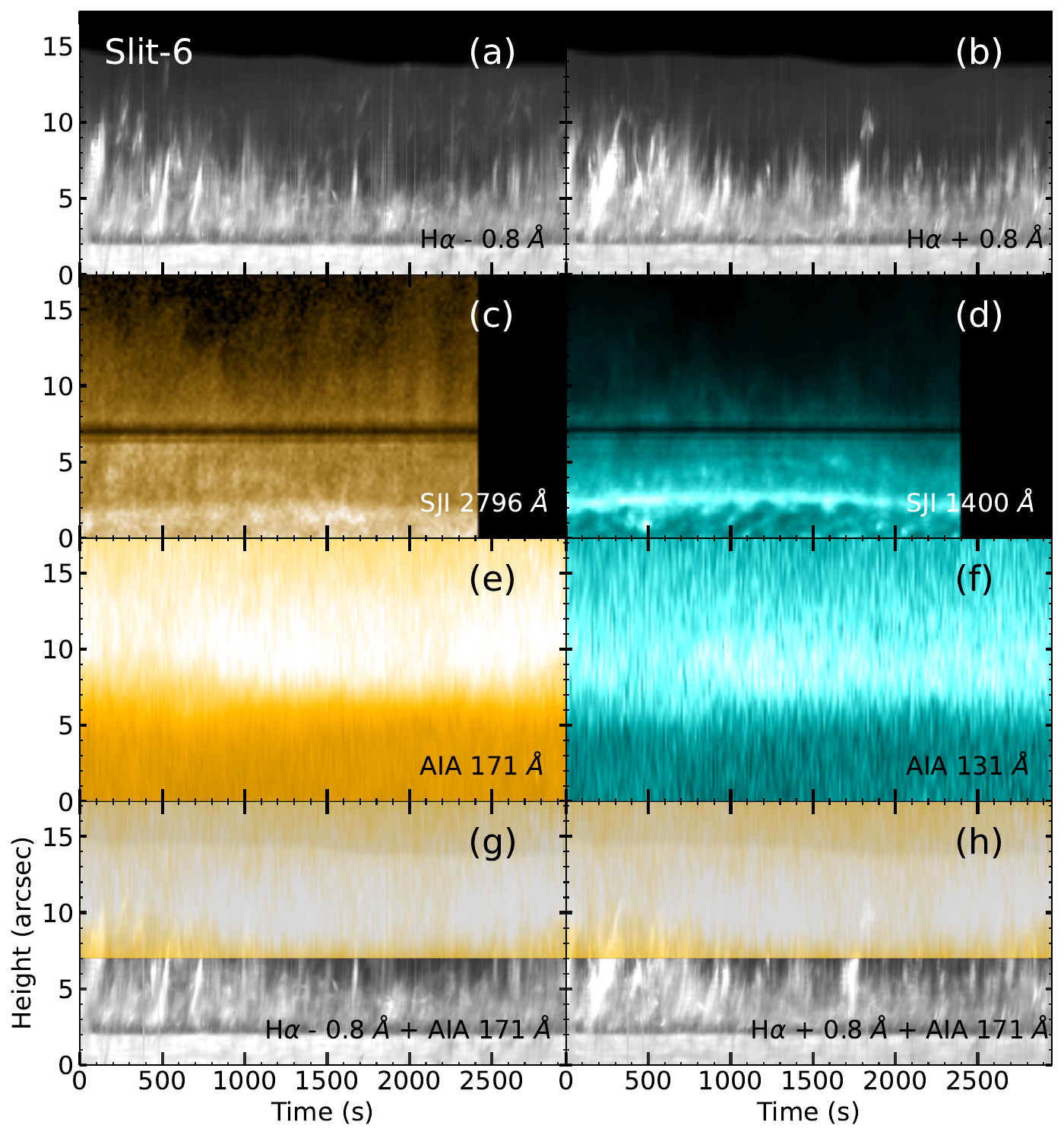}
\caption{Same as Figure~\ref{Space_time_map_slt_1_2_3_4}, but for slit-5 and slit-6. These examples confirm the consistent multi-thermal evolution of the observed jet-like structures.}
\label{Space_time_map_slt_5_6}
\end{figure*}

\begin{figure}
\includegraphics[width=90mm]{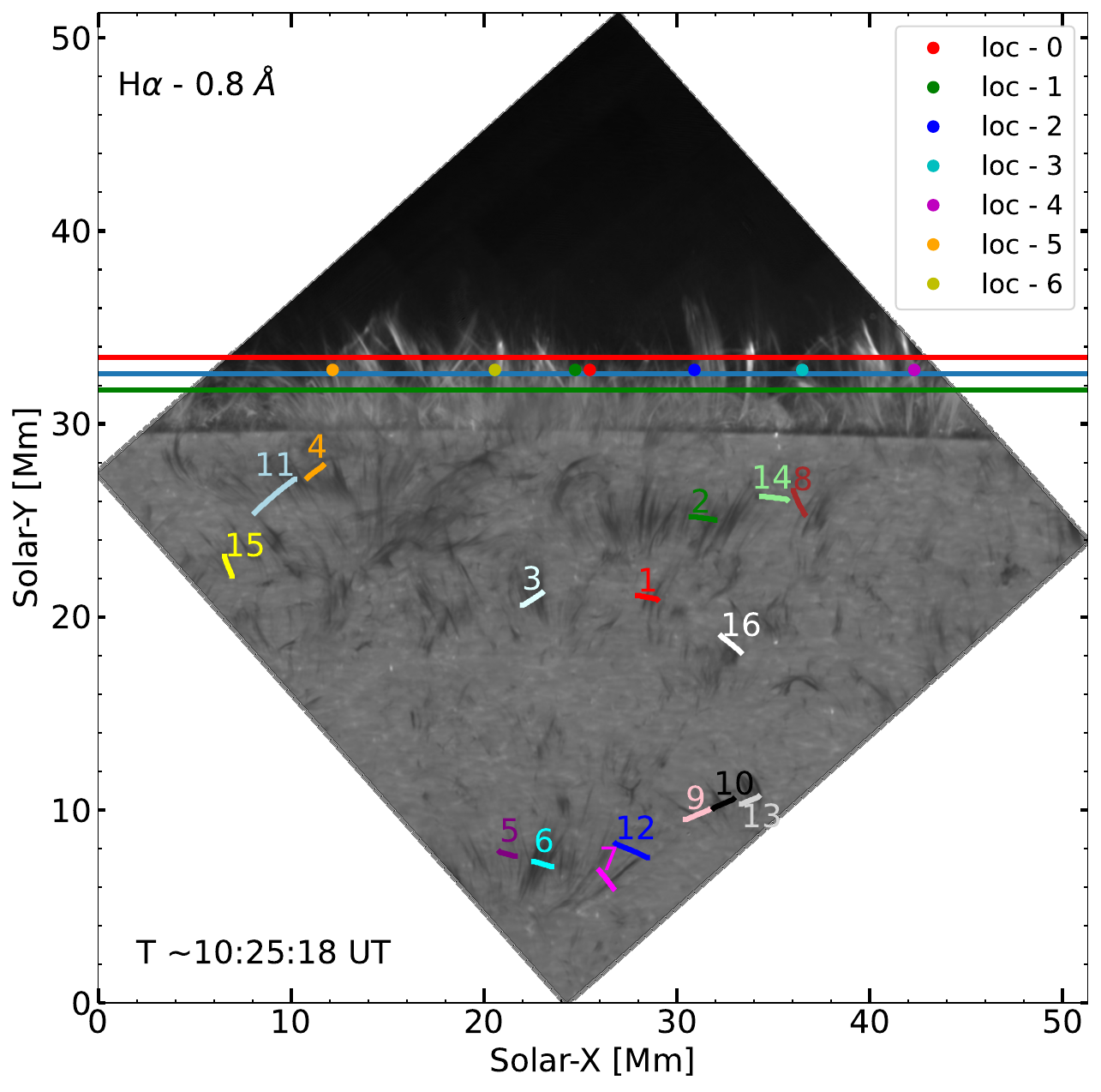}
\caption{Context image showing the locations of the 16 virtual slits used to investigate transverse waves in spicules (numbered and marked by coloured segments), whose space--time diagrams are presented in Figure~\ref{tranverse_wave}. The blue horizontal line indicates the position of the IRIS spectrograph slit. The seven coloured circles mark the selected locations along the IRIS slit where the IRIS spectra are compared with the SST H$\alpha$ spectra (see Figure~\ref{SST_IRIS_spec}) and used for the wavelet analysis (see Figure~\ref{Wavelet}). The green and red horizontal lines indicate positions approximately 0.84~Mm below and above the IRIS slit, respectively. The background image is an SST H$\alpha$ $-0.8$~\AA\ map at the reference time indicated in the lower-left corner.}
\label{slit_loc_wave_SST_IRIS_spec}
\end{figure}

\begin{figure*}
	\includegraphics[width=180mm]{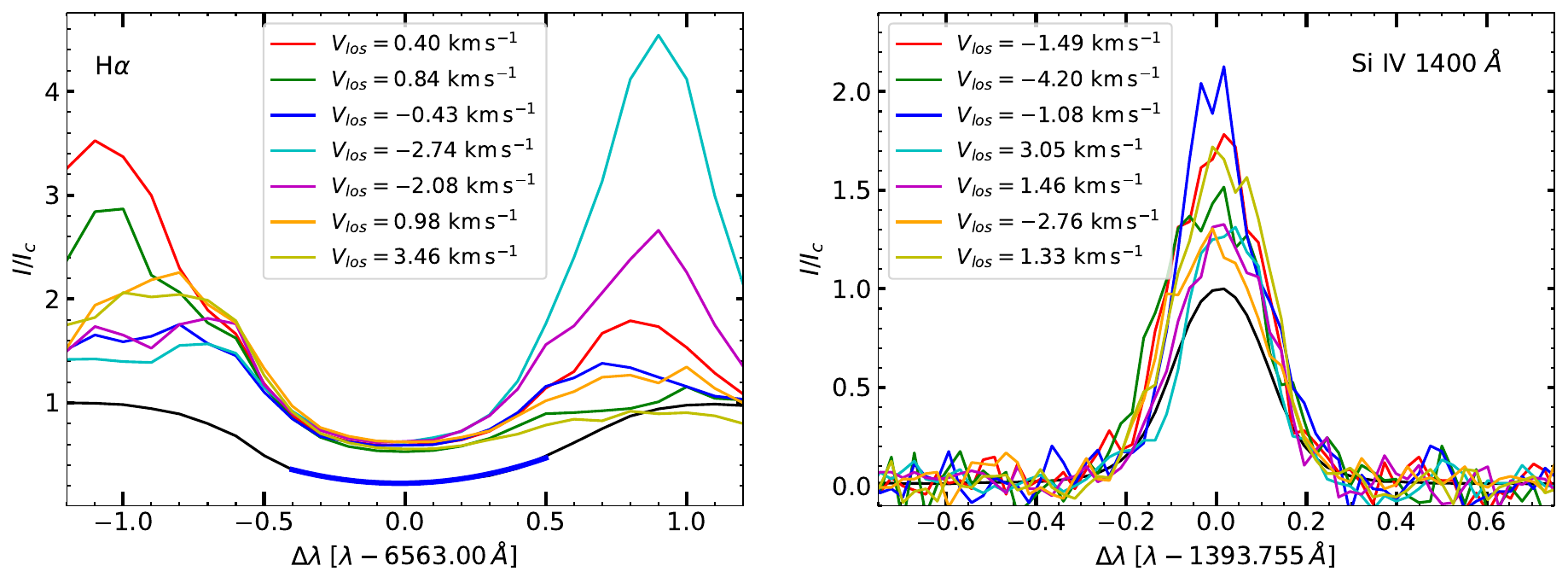}
    \caption{Observed spectral profiles at the seven selected locations marked in Figure~\ref{slit_loc_wave_SST_IRIS_spec}. The left panel shows the SST H$\alpha$ line profiles, while the right panel shows the corresponding IRIS \ion{Si}{iv} 1400~\AA\ line profiles. The black curve in each panel represents the reference average profile used for estimating the relative Doppler shifts. In the left panel, the thick blue curve overplotted on the average H$\alpha$ profile shows the quadratic (parabolic) fit used to determine the line-centre position, where the minimum of the fitted parabola is taken to estimate the Doppler shift of the H$\alpha$ line. The different coloured curves represent the spectra at the seven selected spatial locations, with the corresponding line-of-sight Doppler velocities indicated in the legends.}
	\label{SST_IRIS_spec}
\end{figure*}

\begin{figure*}
	\includegraphics[width=60mm]{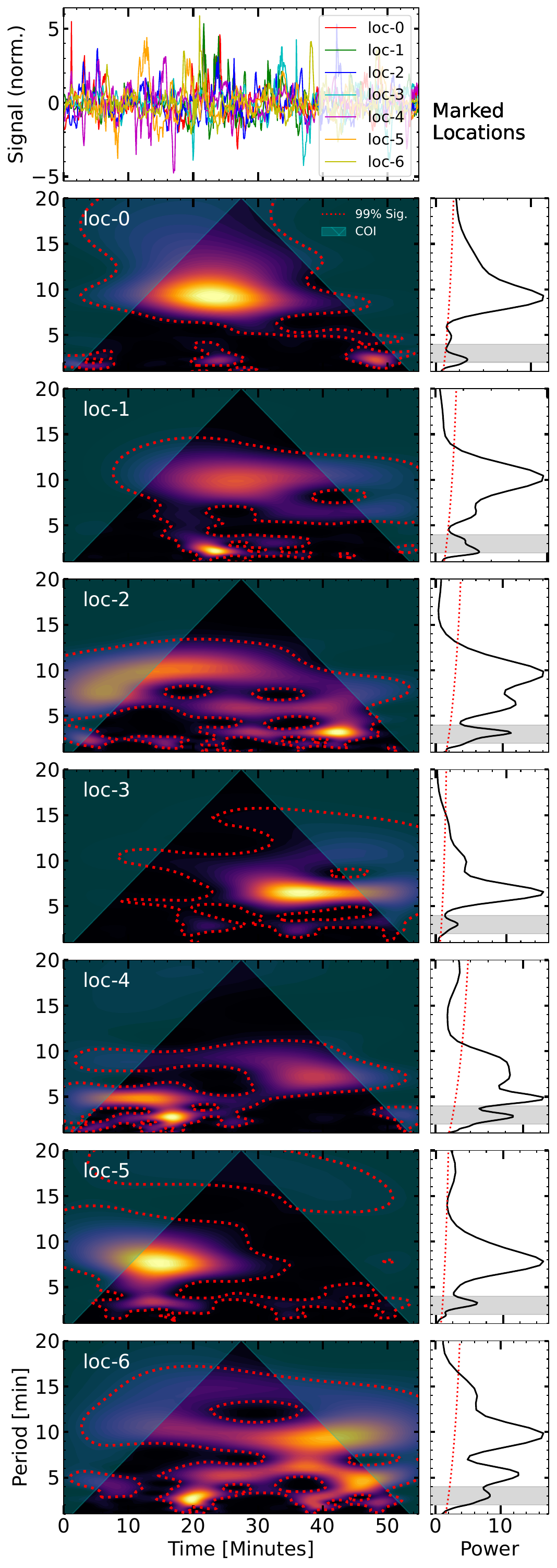} 
	\includegraphics[width=62mm]{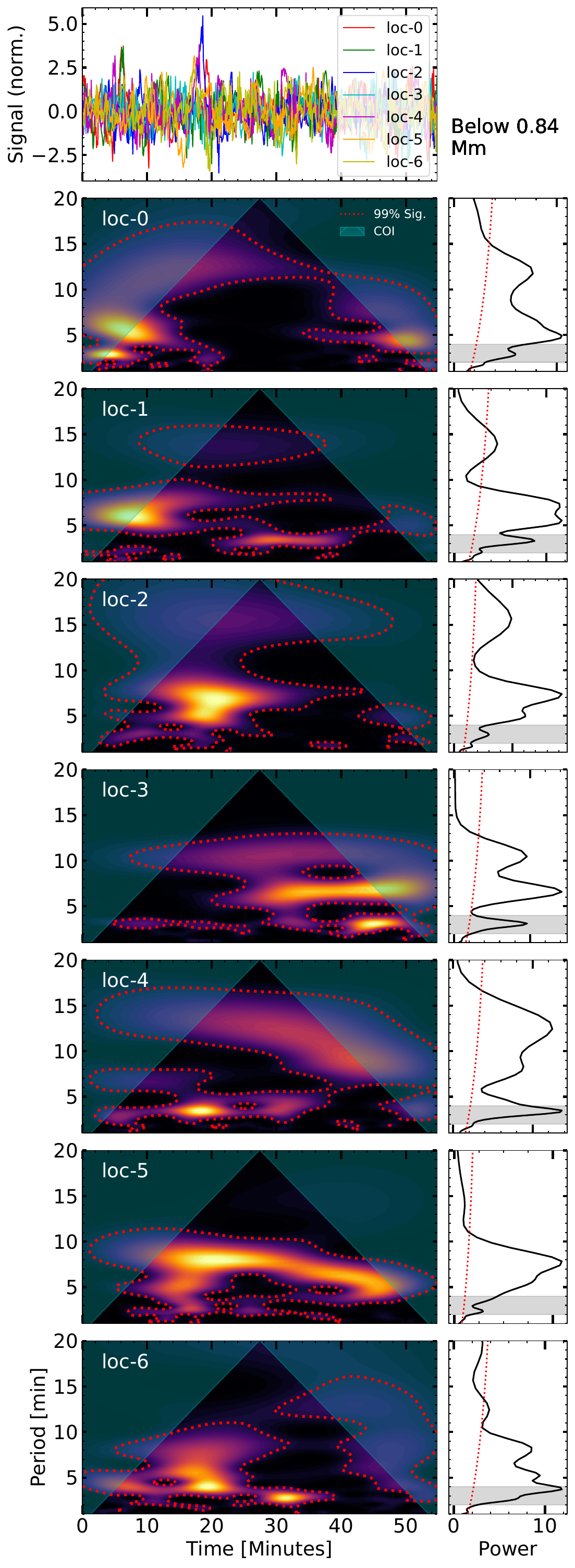}
	\includegraphics[width=63.1mm]{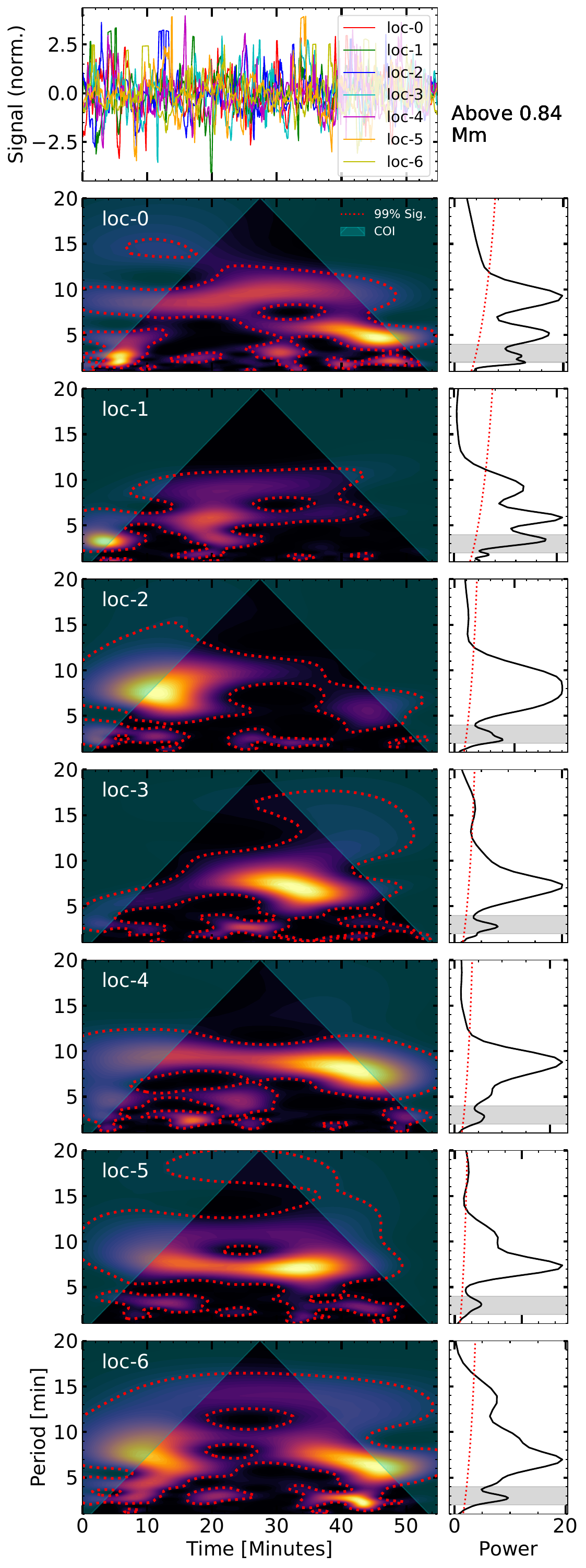}
	\caption{Wavelet analysis of the H$\alpha$ line-centre velocity fluctuations at the seven selected locations shown in Figure~\ref{slit_loc_wave_SST_IRIS_spec}. The left column shows the results at the reference locations along the IRIS slit. The middle and right columns show the corresponding wavelet power spectra at positions approximately 0.84~Mm below and 0.84~Mm above the reference locations, respectively. In each panel, the top sub-panel shows the normalised velocity time series. The colour map represents the wavelet power as a function of time and period, with the red dotted contours indicating the 99\% confidence level. The cross-hatched regions mark the cone of influence (COI). The rightmost sub-panels show the corresponding global wavelet power spectra, with the red dotted line indicating the 99\% significance level. The gray shaded region represents the period range between 2 and 4 minutes.}
	\label{Wavelet}
\end{figure*}

\begin{figure*}
	\includegraphics[width=180mm]{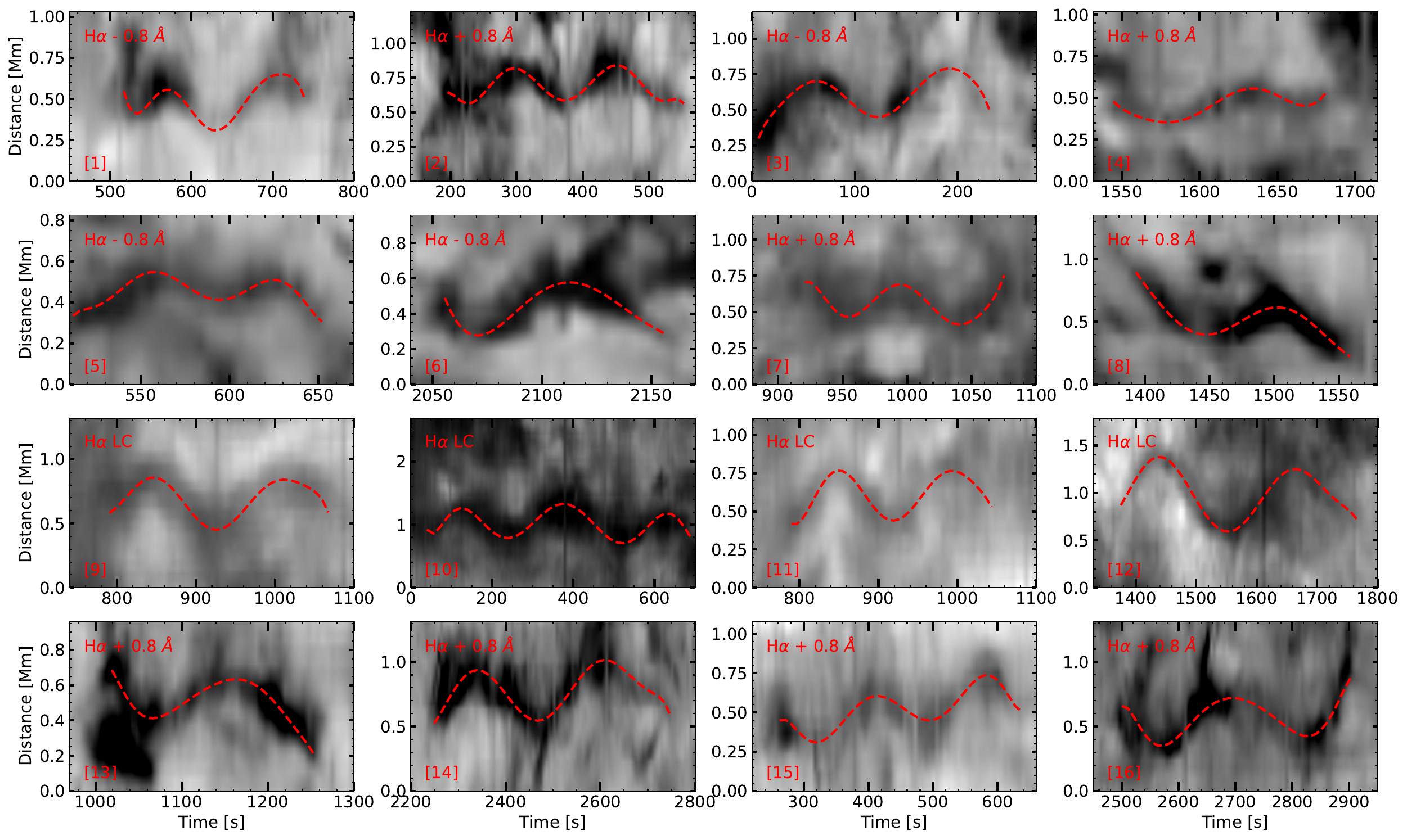}
	\caption{Space--time diagrams for the 16 locations marked in Figure~\ref{slit_loc_wave_SST_IRIS_spec}, illustrating the transverse oscillations of the spicules. The panels show SST H$\alpha$ $-0.8$~\AA, H$\alpha$ line centre (LC), and H$\alpha$ $+0.8$~\AA\ observations, as indicated in each panel. The horizontal axis represents time, and the vertical axis shows the transverse displacement (distance) in Mm. The red dashed curves represent polynomial fits to the transverse motion of the spicule axis. The numbers in the lower-left corner of each panel correspond to the slit identification numbers shown in Figure~\ref{slit_loc_wave_SST_IRIS_spec}. The derived wave parameters are listed in Table \ref{Table_wave_energy}.}
	\label{tranverse_wave}
\end{figure*}

\subsection{Solar Dynamics Observatory (SDO)}

Context observations from the Atmospheric Imaging Assembly (AIA; \citealt{Lemen.et.al.12}) onboard the Solar Dynamics Observatory (SDO; \citealt{Pesnell.et.al.12}) were obtained using the cutout service. We used EUV images from the AIA 171~\AA\ and 193~\AA\ channels, as well as UV images from the 1600~\AA\ and 1700~\AA\ channels. The AIA images have a spatial sampling of $0\farcs6$ per pixel. The EUV channels have a temporal cadence of 12~s, while the UV channels have a cadence of 24~s. To ensure accurate co-alignment, the AIA images from different channels were first aligned with each other. The aligned images were then resampled to match the SST spatial sampling. The final alignment between SDO and SST observations was achieved by cross-correlating nearly simultaneous AIA 1700~\AA\ images with SST H$\alpha$ wing images, which sample similar atmospheric heights.

Figure~\ref{overview} presents an overview of the coordinated observations, showing the chromospheric structures observed in the H$\alpha$ line wing and line centre, together with the co-aligned IRIS SJI 1400~\AA\ and AIA 171~\AA\ images, which sample the transition region and the lower corona, respectively.

\section{Results} \label{Sec3}

\subsection{Dynamical properties of the observed spicules}

We first examine whether the dynamical properties of these spicules are broadly consistent with those reported for spicules in previous studies. In particular, earlier observations have shown that spicules frequently exhibit rapid upward motions, projected velocities of several tens of km s$^{-1}$, and approximately parabolic trajectories associated with decelerating motions.

To assess these characteristics, we constructed space--time diagrams along a selected virtual slit (see Figure \ref{overview}; green slit) and examined shorter temporal windows to better isolate the evolution of individual spicules. Representative examples are shown in Figure \ref{fig:parabolic_path}. The temporal evolution of the spicule apex was approximated using a parabolic fit, while the initial upward phase was fitted with a linear function to estimate the projected upward velocity.

The inferred projected velocities are of the order of ~30--40 km s$^{-1}$, which is broadly consistent with velocity ranges previously reported for limb spicules \citep{DePontieu.et.al.07,Pereira.et.al.12}. Having established that the analyzed structures exhibit characteristic spicular dynamics, we next investigate their multi-thermal evolution across chromospheric, transition-region, and coronal diagnostics.

\subsection{Multi-thermal evolution of spicules}

To investigate the multi-thermal evolution of spicules across different layers of the solar atmosphere, we constructed space–time diagrams using virtual slits placed along selected spicular structures. The slit positions were chosen to sample different regions within the field of view and to follow the orientation of clearly identifiable, elongated spicules. A total of six representative events were analysed. The locations of these virtual slits are marked by cyan lines in Figure~\ref{overview}. Each slit has a width of seven SST pixels, corresponding to a projected width of approximately 0.29~Mm on the plane of the sky. The resulting space--time diagrams are shown in Figure~\ref{Space_time_map_slt_1_2_3_4} for slits 1--4 and in Figure~\ref{Space_time_map_slt_5_6} for slits 5 and 6. These diagrams sample the temporal evolution of spicular structures from the chromosphere, as observed in SST H$\alpha$, to the transition region and corona, as observed in IRIS and AIA channels.

\subsection{Spectral signatures and Doppler velocities}

To quantify the velocity variations associated with the spicules at different atmospheric heights, we analysed the spectral profiles obtained from SST H$\alpha$ and IRIS \ion{Si} {iv} 1394~\AA\ observations at selected locations. These locations are indicated by coloured circles in Figure~\ref{slit_loc_wave_SST_IRIS_spec}. Since the observations were performed near the solar limb, the derived line-of-sight velocities primarily correspond to transverse plasma motions.

The Doppler velocities were estimated by determining the wavelength shift of the spectral line-centres relative to the averaged profile, which served as a reference profile. For the H$\alpha$ line, the line-centre position was obtained by fitting a second-order polynomial (parabola) around the line core, whereas for the \ion{Si} {iv} 1394~\AA\ line, a Gaussian function was fitted to the observed spectral profile. The derived wavelength shifts were subsequently converted into Doppler velocities.

The reference profile for H$\alpha$ was constructed by averaging all spectral profiles over the on-disk field of view for ten consecutive line scans. To assess the robustness of this choice, we also tested alternative averaging procedures, including averaging profiles within a 100 $\times$ 70 pixel quiet-sun region (devoid of obvious spicular structures) over ten consecutive line scans. The resulting reference line-centre positions differed by only $\sim$ 0.02 km s$^{-1}$, indicating negligible sensitivity to the adopted averaging procedure. For the \ion{Si} {iv} 1394~\AA\ line, the reference profile was obtained by averaging spectra over approximately 100 temporal frames and 500 spatial pixels along the slit. The resulting spectra and derived Doppler velocities for both H$\alpha$ and \ion{Si} {iv} 1394~\AA\ are presented in Figure~\ref{SST_IRIS_spec}.

\subsection{Wavelet analysis of chromospheric velocity oscillations}

To investigate the temporal behaviour of velocity oscillations, we performed a wavelet analysis \citep{Torrence&Compo.98} of the H$\alpha$ line-centre velocity time series at same selected locations where the Doppler velocities from H$\alpha$ and \ion{Si} {iv} 1394~\AA\ have been compared (See Figure \ref{slit_loc_wave_SST_IRIS_spec}). In addition to the reference positions, the analysis was also carried out at locations approximately 0.84~Mm above and below each reference point, while keeping the horizontal position fixed. This approach allows us to examine the variation of oscillatory behaviour with height in the chromosphere.

The results of the wavelet analysis are shown in Figure~\ref{Wavelet}, which displays the temporal evolution of the oscillatory power and the corresponding dominant periods. It should be noted that wavelet analysis was not applied to the \ion{Si} {iv} velocity measurements due to the presence of pointing jitter in the IRIS observations.

\subsubsection{Estimation of transverse wave parameters: period, amplitude, velocity, and energy flux}

To quantify the transverse oscillatory motions of individual spicules on disk, we tracked the displacement of the spicule axis as a function of time using the space--time diagrams shown in Figure~\ref{tranverse_wave}. The central ridge of each oscillating structure was identified, and a polynomial function was fitted to trace its temporal evolution. This procedure provides the transverse displacement, $y(t)$, of the spicule axis.

The oscillation period was determined by measuring the time interval between successive intensity maxima in the displacement curves. In cases where only one maximum was clearly identifiable, the period was estimated from the time difference between a complete crest and the adjacent trough.

The displacement amplitude, $A$, was calculated as half of the peak-to-peak variation,

\begin{equation}
	A = \frac{y_{\max} - y_{\min}}{2},
\end{equation}

where $y_{\max}$ and $y_{\min}$ represent the maximum and minimum transverse displacements, respectively.

The transverse velocity was obtained from the temporal derivative of the displacement,

\begin{equation}
	v(t) = \frac{dy}{dt}.
\end{equation}

Since the observed oscillations are not strictly sinusoidal, we estimated the characteristic velocity amplitude using the root-mean-square (RMS) velocity,

\begin{equation}
	v_{\rm rms} =
	\sqrt{
		\left<
		\left(
		\frac{dy}{dt}
		\right)^2
		\right>
	}.
\end{equation}

The energy flux carried by the transverse waves was then estimated using

\begin{equation}
	F =
	\frac{1}{2}
	\rho
	v_{\rm rms}^{2}
	V_A,
\end{equation}

where $\rho$ is the plasma density and $V_A$ is the Alfv\'en speed. We adopted typical chromospheric spicule values of $\rho = 5\times10^{-10}$ kg m$^{-3}$ and $V_A = 200$ km s$^{-1}$ \citep{DePontieu.et.al.07c}.

The measured transverse oscillation parameters and the corresponding energy flux values are summarised in Table~\ref{Table_wave_energy}. The uncertainties in the transverse wave parameters were estimated using standard error propagation. The uncertainty in the displacement amplitude was taken as one SST pixel($\delta$A = 42~Km) while the uncertainty in the period was taken as the temporal cadence of the observations ($\delta$P = 5.5~s).

Since the velocity depends on both amplitude and period, its uncertainty is given by

\begin{equation}
	\delta v =
	v
	\sqrt{
		\left(\frac{\delta A}{A}\right)^2
		+
		\left(\frac{\delta P}{P}\right)^2
	}.
\end{equation}

The uncertainty in the energy flux was estimated by propagating the velocity uncertainty into the energy expression,

\begin{equation}
	\delta F =
	F
	\left(
	2 \frac{\delta v}{v}
	\right).
\end{equation}

\begin{table*}
	\centering
	\caption{Measured transverse oscillation parameters and estimated energy flux carried by the transverse waves. The energy flux is calculated using $F=\frac{1}{2}\rho v^{2} V_{A}$, assuming an Alfv\'en speed $V_{A}=200$ km s$^{-1}$ and spicular density $\rho=5\times10^{-10}$ kg m$^{-3}$ \citep{DePontieu.et.al.07c}. The uncertainties in the mean values are derived from the standard error propagation using the SST pixel scale and temporal cadence.}
	\label{Table_wave_energy}
	\begin{tabular}{ccccc}
		\hline\hline
		   Slit & Period (s) & Amplitude (km) & Velocity (km s$^{-1}$) & Energy flux (W m$^{-2}$) \\
		\hline
		1  & 142 & 236.08 & 7.88 & 3105.16 \\
		2  & 156 & 191.56 & 4.79 & 1149.44 \\
		3  & 126 & 338.37 & 9.85 & 4846.86 \\
		4  & 87  & 140.11 & 6.20 & 1923.69 \\
		5  & 65  & 121.11 & 5.37 & 1441.42 \\
		6  & 83  & 150.22 & 9.91 & 4910.81 \\
		7  & 83  & 170.12 & 8.49 & 3604.97 \\
		8  & 122 & 338.01 & 7.83 & 3062.51 \\
		9  & 163 & 203.41 & 5.56 & 1547.39 \\
		10 & 251 & 312.32 & 5.04 & 1268.90 \\
		11 & 142 & 175.32 & 5.50 & 1510.36 \\
		12 & 230 & 393.86 & 7.10 & 2519.21 \\
		13 & 191 & 239.21 & 4.75 & 1128.26 \\
		14 & 269 & 246.23 & 3.94 & 774.73 \\
		15 & 172 & 213.77 & 3.32 & 551.06 \\
		16 & 257 & 262.40 & 4.25 & 902.08 \\
		\hline
		Mean & 159 & $233 \pm 43$ & $6.23 \pm 1.14$ & $(2.14 \pm 0.78)\times10^{3}$ \\
		\hline
	\end{tabular}
\end{table*}

\section{Discussion}\label{Sec4}

We made use of a unique coordinated dataset between the SST, IRIS, and SDO, which allowed us to trace the evolution of spicules across the chromosphere, transition region, and corona.

Our multi-wavelength space–time analysis reveals a clear correspondence between chromospheric spicules observed in the H$\alpha$ wings and enhanced emission signatures detected in transition region and coronal passbands. The associated coronal brightening is predominantly concentrated near the tip of the spicules and evolves coherently with their motion. This behaviour is clearly illustrated in the composite space--time maps constructed using H$\alpha$ $\pm$ 0.8~\AA\ and AIA 171~\AA\ observations (bottom panels (g) and (h) of Figures~\ref{Space_time_map_slt_1_2_3_4} and \ref{Space_time_map_slt_5_6}). The coronal brightenings follow the extension of spicules: when spicules are shorter, they appear lower, and when spicules are taller they appear higher. This strongly suggests that the enhanced coronal emission is connected to the tops of spicules. This happens not only for a few spicules, but consistently over the entire field of view. To our knowledge, this is the first and most clear detection of spicule coronal signatures seen at the limb.

This observational behaviour is consistent with the on-disk findings reported by \cite{Chaurasiya.et.al.24}, who showed that coronal emission enhancements are co-spatially associated with the tip of spicular structures. If the on-disk scenario described in their study is extended to the limb geometry, the expected observational signature would naturally appear as enhanced emission concentrated near the apparent tip of the spicule, as observed in the present study. In this interpretation, the apparent localisation of coronal brightening near the tip reflects the same physical process viewed from a different line-of-sight perspective, thereby reinforcing the interpretation that spicules are directly associated with coronal plasma heating and emission.

In addition to this geometrical interpretation, another possible physical scenario may also contribute to the observed emission distribution. Spicules are known to exhibit multi-thermal plasma properties (e.g., \citealt{Skogsrud.et.al.15,Chaurasiya.et.al.24}), which is also evident in our space--time maps showing significant simultaneous enhancement in chromospheric and transition-region diagnostics. The presence of cooler and partially ionised plasma within spicules may, in principle, lead to attenuation of background coronal radiation along the line of sight. However, simple estimates indicate that the extinction at 171\,\AA\ is small. For a typical spicule mass density of $\rho \sim 5 \times 10^{-10} \, \mathrm{kg\,m^{-3}}$, corresponding to an electron density of $n_e \sim 3 \times 10^{17} \, \mathrm{m^{-3}}$, and assuming a characteristic spicule width of $L \sim 300 \, \mathrm{km}$, the Thomson scattering optical depth is only $\tau \sim 6 \times 10^{-6}$, indicating that continuum attenuation due to Thomson scattering is negligible. 

In addition, bound-free (photoionization) absorption at 171\,\AA\ can arise from \ion{H} {i}, \ion{He} {i}, and \ion {He} {ii}. Estimates of the total extinction coefficient show that the dominant contribution arises from He\,\textsc{i} bound-free opacity at lower temperatures; however, even under conditions representative of dense spicular plasma, the resulting extinction remains small. Moreover, the opacity decreases rapidly above $\sim 2 \times 10^{4}$~K as He\,\textsc{i} becomes ionised. For typical values of the extinction coefficient ($\alpha \sim 10^{-7} \, \mathrm{m^{-1}}$) and path lengths of $L \sim 300 \, \mathrm{km}$, the corresponding optical depth is $\tau \sim 0.03$, which remains modest. Therefore, even when considering multiple spicules along the line of sight in limb observations, it is difficult to achieve sufficiently large optical depths to significantly attenuate the background 171\,\AA\ emission.

The coordinated spectroscopic analysis of SST H$\alpha$ and IRIS \ion{Si} {iv} 1394~\AA\ observations reveals complex plasma dynamics within the multi-thermal spicular structures. The line-of-sight velocity measurements show that chromospheric and transition-region plasma components do not always evolve coherently and may exhibit opposite Doppler signatures at the same projected location and nearly the same time. Since the observations were obtained close to the solar limb, the measured LOS velocities are expected to predominantly reflect transverse plasma motions rather than purely field-aligned flows.

The physical origin of these temperature-dependent Doppler signatures remains uncertain. One possible explanation is that unresolved multi-stranded structures along the LOS contribute differently to the H$\alpha$ and \ion{Si} {iv} 1394~\AA\ signals, particularly because H$\alpha$ is optically thick while \ion{Si} {iv} 1394~\AA\ is optically thin. Alternatively, torsional or rotational motions within multi-thermal spicular structures may cause different temperature diagnostics to preferentially sample different regions of the same structure. We also cannot exclude the possibility that thermally distinct plasma components evolve differently within the same projected magnetic structure. Therefore, while the observations clearly indicate dynamically complex multi-thermal behaviour, the precise physical mechanism responsible for the observed opposite Doppler signatures cannot be uniquely determined from the present data.

To investigate the oscillatory behaviour associated with spicules, we performed wavelet analysis of horizontal velocity signals at different heights. Our results show that oscillations with periods around $\sim$3~minutes are commonly present at multiple locations and heights, exhibiting significant power above the 99\% confidence level. These oscillations are consistent with chromospheric oscillations frequently observed in the solar atmosphere (with the vertical velocity fluctuations) and are likely associated with magnetoacoustic wave propagation along magnetic structures \citep{Jess.et.al.15,Jess.et.al.23}. Furthermore, the wavelet power distribution shows that oscillatory power is not uniformly distributed but is often concentrated at specific times. The intermittent nature of the wave power observed in our wavelet analysis suggests that wave excitation may be episodic rather than continuous. This behaviour may be related to impulsive energy release events, magnetic reconnection, or buffeting of magnetic flux tubes by convective motions in the photosphere. The presence of these oscillations across different heights indicates efficient wave propagation within spicules.

In addition to the longer-period oscillations discussed above, we detect clear signatures of high-frequency transverse waves within individual spicules. The space–time analysis of the transverse displacement reveals oscillation periods ranging from 65 to 269 s, with a mean period of 159 s. The corresponding displacement amplitudes vary between 121 and 394 km, with a mean value of $233 \pm 43$ km, while the transverse velocity amplitudes range from 3.3 to 9.9 km s$^{-1}$, with a mean value of $6.23 \pm 1.14$ km s$^{-1}$. Using these velocity measurements, we estimate the associated energy flux to lie between $5.5\times10^{2}$ and $4.9\times10^{3}$ W m$^{-2}$, with a mean value of $(2.14 \pm 0.78)\times10^{3}$ W m$^{-2}$. The detection of these transverse waves is enabled by the high spatial and temporal resolution of the SST observations, which allow precise tracking of the spicule axis. These oscillations manifest as periodic lateral displacements of the spicule structure and are consistent with magnetohydrodynamic (MHD) kink-mode waves propagating along magnetic flux tubes \citep{Zaqarashvili&Erdelyi.09}.

Transverse oscillations of this nature have been reported in numerous previous studies. Using Hinode observations, \citet{DePontieu.et.al.07c} first identified transverse waves in spicules and suggested that they may carry sufficient energy to contribute to coronal heating. Similarly, \citet{Okamoto&DePontieu.11} detected transverse motions in chromospheric structures and interpreted them as Alfv\'enic waves. Such waves are expected to arise naturally in thin magnetic flux tubes embedded within a magnetised plasma and can efficiently transport energy upward along magnetic field lines.

Our measured wave properties are broadly consistent with these earlier observational results, although some important differences are present. For example, \citet{DePontieu.et.al.07c} reported periods between 100 and 500 s, velocity amplitudes of 10–30 km s$^{-1}$, and displacement amplitudes of 200–500 km. While our displacement amplitudes are comparable, the measured velocity amplitudes are somewhat lower. However, our observations reveal significantly shorter periods, extending down to 65 s, indicating the presence of higher-frequency wave components that were not as clearly resolved in earlier studies. These differences likely reflect the improved spatial and temporal resolution of the SST, as well as intrinsic variability among individual spicules and differences in observing conditions.

The estimated energy flux in our study (derived primarily from spicule-dominated strands structures), with a mean value of $(2.14 \pm~0.78)$ $\times$ 10$^{3}$~Wm$^{-2}$, substantially exceeds the typical energy requirement for heating the quiet solar corona ($\sim$100--200 W~m$^{-2}$; \citealt{Withbroe&Noyes.77}). However, this comparison does not include the dissipation height, damping efficiency, and conversion of wave energy into thermal energy. Additionally, we do not explicitly quantify losses due to reflection and mode conversion. Nevertheless, the estimated wave energy flux suggests that these oscillations may contribute significantly to energy transport into the upper atmosphere. However, the role of spicules in coronal heating remains debated. Several observational and numerical studies have argued that spicules provide only a limited contribution to sustaining the hot corona, as direct plasma injection by spicules cannot account for the majority of coronal emission \citep{Klimchuk.12,Tripathi&Klimchuk.13, Patsourakos.et.al.14,Klimchuk&Bradshaw.14, Bradshaw&Klimchuk.15,Sow.Mondal.et.al.22}. Collectively, these studies suggest that while spicules may not be the primary source of coronal plasma or heating, they may still contribute indirectly through waves, shocks, or magnetic energy transport. Importantly, many of these studies focus on direct plasma supply or chromospheric heating scenarios, whereas our estimates concern energy transported by transverse wave motions. Therefore, our estimated wave energy flux should be interpreted as evidence that transverse oscillations in spicules may represent a potentially important channel for energy transport into the upper atmosphere, while their exact contribution to coronal heating remains uncertain.

\section{Conclusion} \label{Sec5}

In this work, we investigated the multi-thermal and dynamical behaviour of solar spicules using coordinated SST/H$\alpha$, IRIS \ion{Si}{iv}, and SDO/AIA observations. Our analysis reveals a clear temporal and spatial correspondence between chromospheric spicules and coronal emission at the limb, indicating a direct connection between plasma at different atmospheric layers. The Doppler velocity measurements can show opposite flow signatures in H$\alpha$ and \ion{Si}{iv}, providing evidence for the coexistence of distinct multi-thermal plasma components within individual spicules. We also find that high-frequency transverse oscillations in spicules (periods of 65–270 s) can provide more than sufficient energy flux to contribute to coronal heating. In addition, we identify frequently dominant $\sim$3-minute oscillations across multiple heights, highlighting the presence of coherent chromospheric oscillatory processes within these structures. 

Our findings demonstrate that spicules are not only multi-thermal and highly dynamic, but also can be efficient channels for both mass and wave energy transport into the upper solar atmosphere. The combined evidence of multi-thermal plasma flows and energetically significant transverse waves highlights their important role in coupling the chromosphere to the transition region and corona, and in contributing to the overall energy balance of the solar atmosphere.

\begin{acknowledgements}

TMDPs work has been supported by the Research Council of Norway through its Centres of Excellence scheme, project number 262622. The Swedish 1-m Solar Telescope is operated on the island of La Palma by the Institute for Solar Physics of Stockholm University in the Spanish Observatorio del Roque de los Muchachos of the Instituto de Astrofísica de Canarias. The Swedish 1-m Solar Telescope, SST, is co-funded by the Swedish Research Council as a national research infrastructure (registration number 4.3-2021-00169). SDO is a mission for NASA Living With a Star program. The SDO/HMI data were provided by the Joint Science Operation Centre (JSOC). IRIS is a NASA small explorer mission developed and operated by LMSAL, with mission operations executed at NASA Ames Research Center and major contributions to downlink communications funded by the ESA and the Norwegian Space Center. We have made use of NASA's Astrophysics Data System Bibliographic Services.

\end{acknowledgements}


\bibliographystyle{aa} 
\bibliography{bibfile}
\end{document}